\documentclass[preprint,showpacs,preprintnumbers,amsmath,amssymb, showkeys,superscriptaddress,titlepage]{revtex4-2}
\usepackage{graphicx}
\usepackage{dcolumn}
\usepackage{bm}
\usepackage{hyperref}
\usepackage{textcomp}

\begin{document}
	
	\title{Dissociation of Relativistic $^{10}$B Nuclei in Nuclear Track Emulsion}
	
	\author{A.A. Zaitsev}
	\email{zaicev@jinr.ru}
	\affiliation{Joint Institute for Nuclear Research (JINR), Dubna, Russia}
	\affiliation{P.N. Lebedev Physical Institute of the Russian Academy of Sciences (LPI), Moscow, Russia}
	
	\author{D.A. Artemenkov}
	\affiliation{Joint Institute for Nuclear Research (JINR), Dubna, Russia}
	
	\author{V. Bradnova}
	\affiliation{Joint Institute for Nuclear Research (JINR), Dubna, Russia}
	
	\author{P.I. Zarubin}
	\affiliation{Joint Institute for Nuclear Research (JINR), Dubna, Russia}
	\affiliation{P.N. Lebedev Physical Institute of the Russian Academy of Sciences (LPI), Moscow, Russia}
	
	\author{I.G. Zarubina}
	\affiliation{Joint Institute for Nuclear Research (JINR), Dubna, Russia}
	
	\author{R. R. Kattabekov}
	\affiliation{Joint Institute for Nuclear Research (JINR), Dubna, Russia}
	
	\author{N.K. Kornegrutsa}
	\affiliation{Joint Institute for Nuclear Research (JINR), Dubna, Russia}
	
	\author{K. Z. Mamatkulov}
	\affiliation{Joint Institute for Nuclear Research (JINR), Dubna, Russia}
	
	\author{E. K. Mitsova}
	\affiliation{Joint Institute for Nuclear Research (JINR), Dubna, Russia}
	\affiliation{South-Western University, Blagoevgrad, Bulgaria}
	
	\author{A. Neagu}
	\affiliation{Institute of Space Science, Magurele, Romania}
	
	\author{P. A. Rukoyatkin}
	\affiliation{Institute of Space Science, Magurele, Romania}
	
	\author{V.V. Rusakova}
	\affiliation{Joint Institute for Nuclear Research (JINR), Dubna, Russia}
	
	\author{V. R. Sarkisyan}
	\affiliation{Yerevan Physics Institute, Yerevan, Armenia}
	
	\author{R. Stanoeva}
	\affiliation{South-Western University, Blagoevgrad, Bulgaria}
		
	\author{M. Haiduc}
    \affiliation{Institute of Space Science, Magurele, Romania}
	
	\author{E. Firu}
	\affiliation{Institute of Space Science, Magurele, Romania}

	\begin{abstract}
		The structural features of $^{10}$B are studied by analyzing the dissociation of nuclei of this isotope at
		an energy of 1 A GeV in nuclear track emulsion. The fraction of the $^{10}$B $\to$ 2He + H channel in the charge
		state distribution of fragments is 78\%. It was determined based on the measurements of fragment emission
		angles that unstable $^{8}$Be$_{g.s.}$ nuclei appear with a probability of (26 $\pm$ 4)\%, and (14 $\pm$ 3)\% of them are produced
		in decays of an unstable $^9$B$_{g.s.}$ nucleus. The Be + H channel was suppressed to approximately 1\%.
	\end{abstract}
	
	\maketitle
	

Virtual nucleon associations (clusters) are the fundamental
structural elements of atomic nuclei. Their
simplest observable manifestations are the lightest
$^{4,3}$He and $^{3,2}$H nuclei, which have no excited states.
Superpositions of the lightest clusters and nucleons
form subsequent nuclei (including unstable $^8$Be and
$^9$B), which act as constituent clusters themselves. The
balance of possible superpositions in states with suitable
spin and parity values defines binding and the
parameters of the ground state of the corresponding
nucleus. Clusterization of the ground state of a light
nucleus defines the structure of its excitations and the
initial conditions of reactions it is involved in. Further
attachment of nucleons and lightest nuclei leads to a
shell-type structure. The entanglement of cluster and
shell degrees of freedom turns the group of light nuclei
into a ``laboratory'' of nuclear quantum mechanics.
Clusterization forms the basis of processes that
accompany the phenomena of physics of nuclear isobars,
hypernuclei, and quark-parton degrees of freedom.
The concept of clusterization of nuclei is essential
to applications in nuclear astrophysics, cosmic-ray
physics, nuclear medicine, and, possibly, nuclear
geology.

The BECQUEREL project \cite{1}, which is focused
on examining the cluster structure of light nuclei,
involved irradiation of nuclear track emulsion (NTE)
with relativistic Be, B, C, and N isotopes (including
radioactive ones) at the JINR Nuclotron \cite{2}. Longitudinally
irradiated NTE layers provide an opportunity to analyze the fragment ensembles fully. The events of
coherent dissociation of nuclei with no tracks of slow
fragments and charged mesons (``white'' stars; see
Fig. 1) are especially valuable in this respect. The irradiation
of NTE with $^{10}$B nuclei with an energy of
1 A GeV was performed in 2002 in the first run at the
extracted Nuclotron beam. The success of this experiment
paved the way for subsequent irradiations with
secondary beams enriched in $^8$B and $^9$Be nuclei that
were formed based on acceleration and fragmentation
of $^{10}$B. The effect of dominance of 2He + H white stars
($\sim$70\%) in the dissociation of $^{10}$B nuclei was noted, but
was not analyzed. In addition, the Be + H channel
turned out to be suppressed (no more than 2\%). This
irradiation was ``overshadowed'' by irradiations with
relativistic radioactive neutron-deficient nuclei. The
discovery of a considerable contribution of an unstable
$^9$B nucleus to the structure of a radioactive $^{10}$C nucleus
\cite{3} highlighted the importance of in-depth analysis of
dissociation $^{10}$B $\to$ 2He + H. This analysis is aimed at
determining the probabilities of coherent dissociation
of a $^{10}$B nucleus involving $^8$Be and $^9$B nuclei. The
continuation of studies into the $^{10}$B nucleus structure
was relevant to interpreting the data on $^{11}$C, where $^{10}$B
may serve as a structural element \cite{4}.

\begin{figure}[h]
	\centerline{\includegraphics[width=16cm]{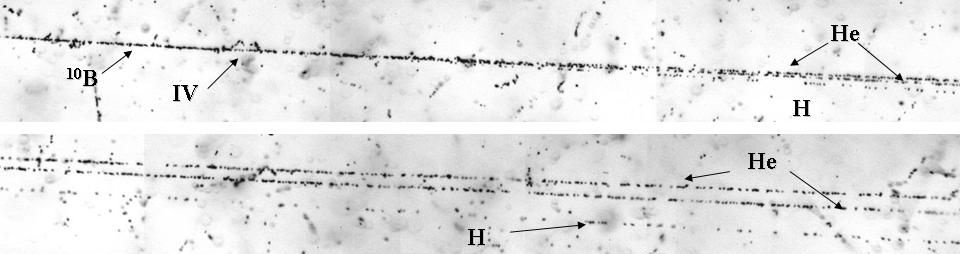}}
	\caption{Macrophotograph of the event of coherent dissociation of a $^{10}$B nucleus into He and H fragments (IV is the approximate
		position of the interaction vertex). This event has the following parameters: $\Theta_{2\alpha}$ = 5.3 mrad, $Q_{2\alpha}$ = 87 keV, and $Q_{2\alpha p}$ = 352 keV.}
\end{figure}

\begin{figure}[h]
	\centerline{\includegraphics[width=15cm]{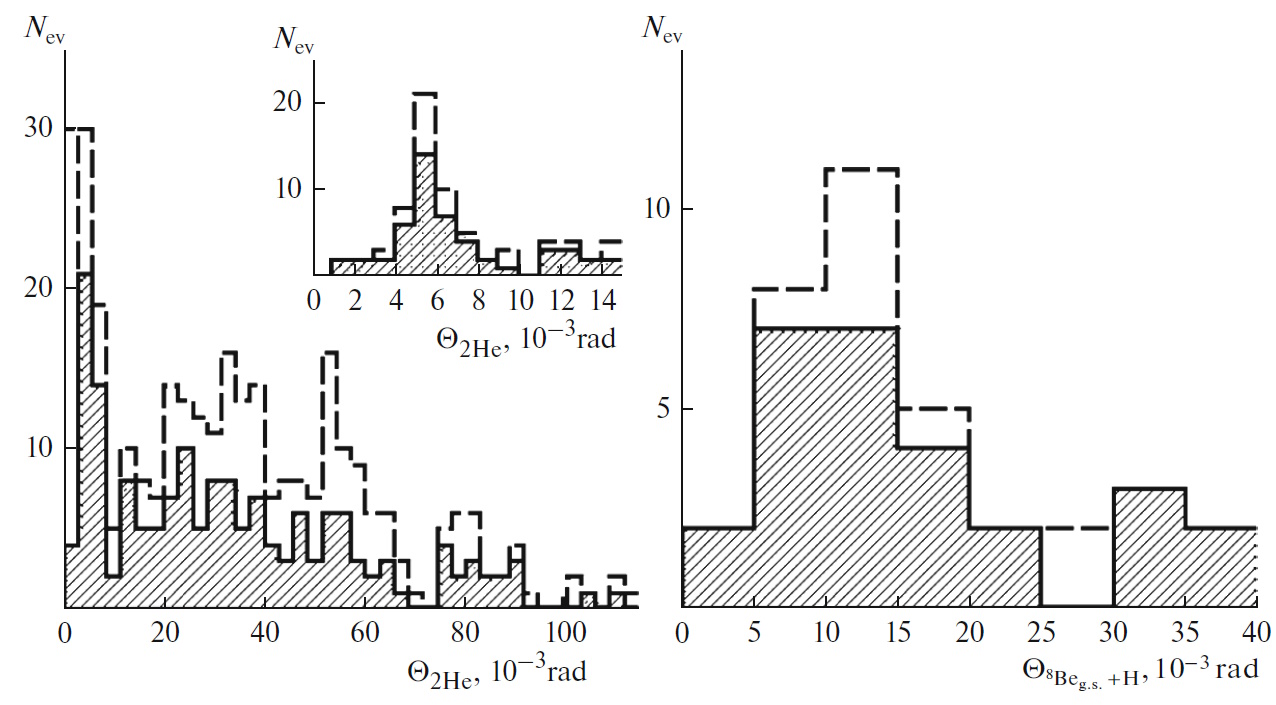}}
	\caption{Distribution of $^{10}$B $\to$ 2He + H events as a function of the opening angle $\Theta_{\textrm{2He}}$ in 2He pairs (left) and the opening angle $\Theta_{^8\textrm{Be+H}}$ in $^8$Be$_{g.s.}$ and H pairs (right) for all found events (dashed curve) and in ``white'' stars (hatched).}
\end{figure}

Nuclei with a marked cluster structure should act
as cores in $^{10}$B. This is evidenced by the thresholds of
separation of nucleons and the lightest nuclei $^6$Li + $\alpha$
(4.5 MeV), $^8$Be + $d$ (6.0 MeV), $^9$Be + $p$ (6.6 MeV), and
$^9$B + $n$ (8.4 MeV). As in the case of $^{10}$C, decays of an
unstable $^9$B nucleus may serve as sources of $^8$Be$_{g.s.}$
nuclei in the ground state 0$^+$ in the process of dissociation
of $^{10}$B. The $^8$Be$_{2^+}$ + $d$ cluster configuration could
serve as the source of $^8$Be nuclei in the first excited
state 2$^+$. Another $^{10}$B component is based on a $^9$Be
nucleus with $^8$Be$_{g.s.}$ and $^8$Be$_{2^+}$ featuring in almost equal
measure in its structure. This component may manifest
itself in the dissociation of $^{10}$B both in the production
of $^9$Be nuclei and in the emergence of pairs of
$\alpha$-particles $^8$Be$_{g.s.}$ and $^8$Be$_{2^+}$. The probability of coherent
dissociation in the $^9$B + $n$ channel is expected to be
the same as that for the $^9$Be + $p$ mirror channel. Likewise,
a $^6$Li nucleus can be present both as an integral
formation and as virtual $\alpha$ + $d$ bonding.

These considerations motivated us to resume the
analysis of $^{10}$B irradiation in 2015. The tracks of beam
$^{10}$B nuclei in NTE have already been examined over
the length of 241 m. Altogether, 1664 inelastic interactions
were found as a result. The charge topology distribution
of 127 $^{10}$B white stars (see Table 1) confirms
the dominance of the 2He + H (78\%) channel and the
suppression of the Be + H (1\%) channel, which
should correspond to the $^9$Be + $p$ configuration.

\begin{table}[h]
	\caption{Distribution of 127 $^{10}$B ``white'' stars over the
		charge configurations of fragments}
\begin{tabular}{c|c}
	\hline
	Be + H & 1 (1\%) \\
	2He + H, including $^8$Be and $^9$B & 99, 24, 13 (78, 19, 10)\% \\
	He + H & 16 (12\%) \\
	Li + He & 5 (4\%) \\
	Li + 2H & 5 (4\%) \\
	5H & 1 (1\%) \\ \hline
\end{tabular}
\end{table}

In order to obtain a reliable reference $^8$Be and $^9$B
signal based on angular measurements, the statistics of
$^{10}$B $\to$ 2He + H events was brought up to 296 (including
166 ``white'' stars). This increase was achieved by
implementing a quick area search and adding ``nonwhite''
2He + H stars to the measurements. The sampling
is governed primarily by the geometric pattern
of events in the emulsion volume relative to the fiducials
and does not introduce any additional selection
criteria.

\begin{figure}[h]
	\centerline{\includegraphics[width=15cm]{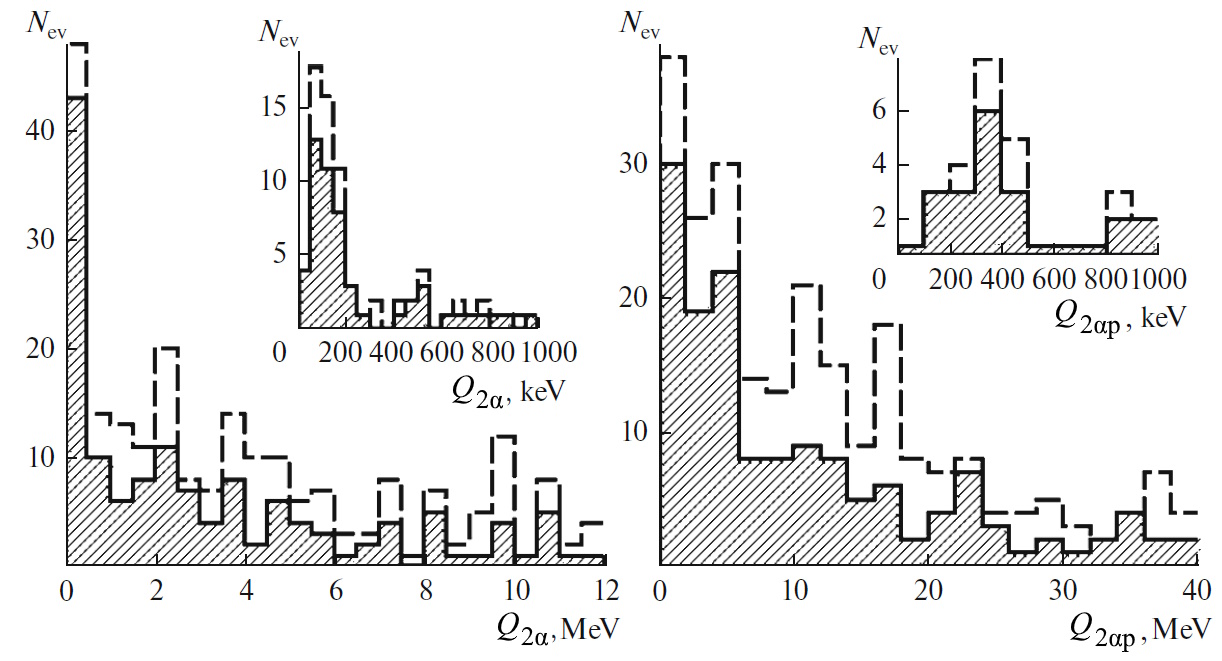}}
	\caption{Distribution of $^{10}$B $\to$ 2He + H events as a function of energy $Q_{2\alpha}$ of $\alpha$-particle pairs (left) and energy $Q_{2\alpha p}$ of 2$\alpha$ + $p$ triples
		(right) for all found events (the dashed curve) and in ``white'' stars (hatched). Magnified portions of distributions over $Q_{2\alpha}$ and $Q_{2\alpha p}$	are shown in the insets.}
\end{figure}

\begin{figure}[h]
	\centerline{\includegraphics[width=8cm]{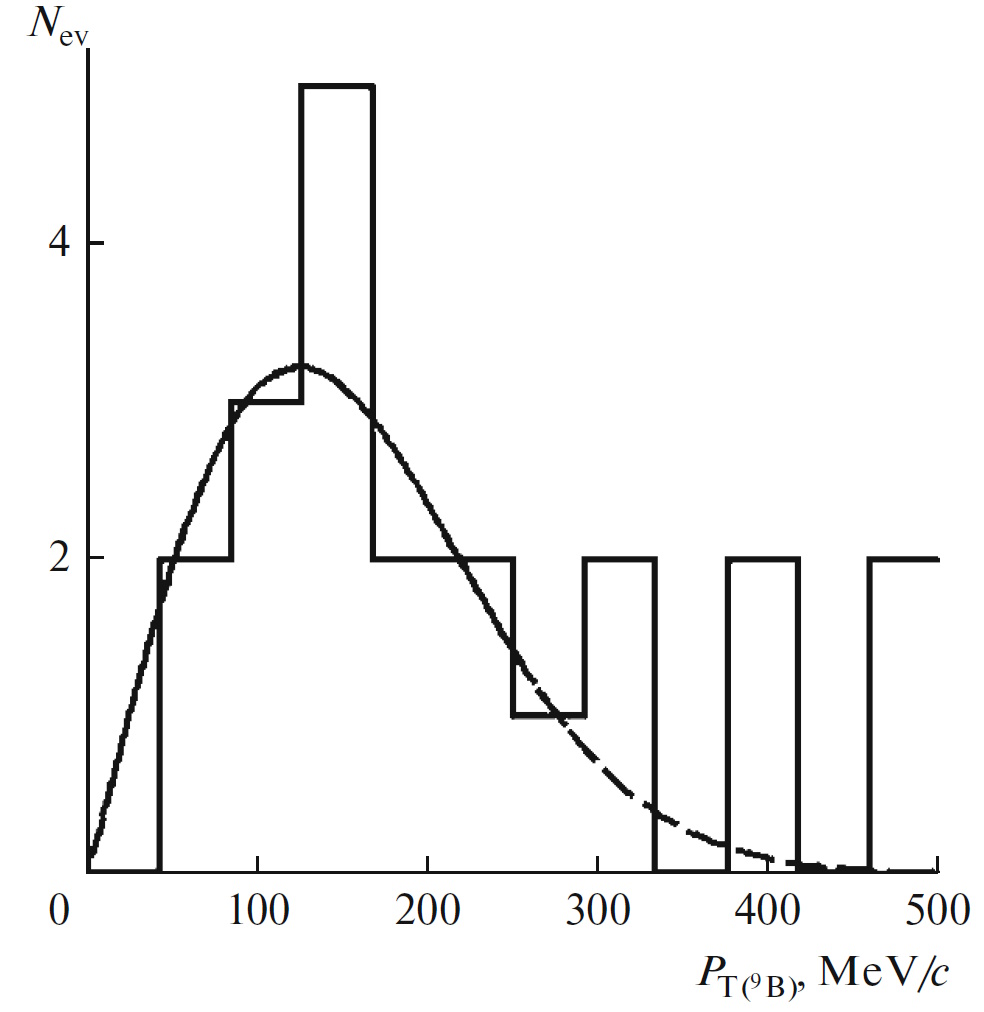}}
	\caption{Distribution of ``white'' stars as a function of the
		transverse momentum $P_\textrm{T}$($^9$B) of 2He + H triples with the
		formation of a $^9$B$_{g.s.}$ nucleus.}
\end{figure}

The distribution of 2He pairs in this sample over
spatial angle $\Theta_{2\textrm{He}}$ (Fig. 2, left panel) in the 0 $<$ $\Theta_{2\textrm{He}}$ $<$
10.5 mrad interval indicates the existence of 56 $^8$Be$_{g.s.}$
decays (with 40 of them from white stars). A total of
28 decays (including 22 decays in white stars; Fig. 2,
right panel) assigned to $^9$B$_{g.s.}$ (as in \cite{5}) can be isolated
from the distribution of all measured events over spatial
angle $\Theta_{2\text{HeH}}$ between the $^8$Be$_{g.s.}$ and H directions in
the 0 $<$ $\Theta_{^8\textrm{Be+H}}$ $<$ 25 mrad interval. Thus, $^8$Be$_{g.s.}$ nuclei
form via $^9$B$_{g.s.}$ decays in just a half of all events.

The decays of relativistic $^8$Be and $^9$B nuclei can be
reconstructed based on excitation energy $Q = M^* – M$,
which is the difference between invariant mass of fragments
$M^*$, $M^{*2} = \Sigma(P_i\cdot P_k)$, and total fragment mass $M$.
$P_{i,k}$ are 4-momenta determined in the approximation
of conservation of the initial momentum of fragments
per nucleon. In the region of small opening angles, it
is reasonable to assume that the H isotope corresponds
to protons and the He isotope corresponds to $\alpha$-particles.
The distribution over energy $Q_{2\alpha}$ (Fig. 3, left
panel) at 0 $<$ $Q_{2\alpha}$ $<$ 200 keV has a mean value of 105 $\pm$
7 keV with RMS = 46 keV and corresponds to the
$^8$Be$_{g.s.}$ ground state, while the distribution over energy
$Q_{2\alpha p}$ of 2$\alpha$ + $p$ triples (Fig. 3, right panel) at 0 $<$ $Q_{2\alpha p}$ $<$
400 keV has a mean value of 261 $\pm$ 23 keV with RMS =
91 keV and corresponds to the $^9$B$_{g.s.}$ ground state. The
distribution over transverse momentum $P_\textrm{T}$($^9$B) of $^9$B$_{g.s.}$
nuclei (Fig. 4) is a Rayleigh one with parameter
$\sigma_{P_\textrm{T}}$($^9$B) = 121 $\pm$ 30 MeV/$c$, which does not contradict
the statistical model (96 MeV/$c$). The He and H isotopes
are now being identified by the multiple scattering
method, which should extend the region of the fragment
opening angles under investigation.

Thus, unstable $^8$Be and $^9$B nuclei manifest themselves
in coherent dissociation in the $^{10}$B $\to$ 2He + H
channel with a probability of (26 $\pm$ 4)\% and (14 $\pm$ 3)\%,
respectively. Therefore, they are significant constituents
of a $^{10}$B nucleus. It is unexpected that the number
of $^9$B + $n$ white stars is ten times higher than that of
$^9$Be + $p$ (see Table 1). This observation may indicate
that the spatial distribution of neutrons in a $^{10}$B
nucleus is wider than that of protons, which results in
a larger cross section of the $^9$B + $n$ channel.

It appears that the physics behind this is as follows.
The $^9$B nucleus is a ``loose'' nuclear-molecular structure
made of 2$\alpha$ + $p$ clusters. The Coulomb barrier can
enhance the proton binding. The $^9$Be core nucleus is
also likely to be present in $^{10}$B not as an integral formation,
but in a ``loose'' form of 2$\alpha$ + $n$ (an approximately
even superposition of $^8$Be$_{g.s.}$ and $^8$Be$_{2^+}$ couplings
with a neutron). The dominance of decays of
$^8$Be$_{g.s.}$ over $^9$B$_{g.s.}$ in the dissociation may be attributed
to the additional contribution of a ``loose'' $^9$Be
nucleus. Note that $^9$Be may not be present in the
structure of $^{10}$C. Indeed, the decays of $^8$Be$_{g.s.}$ in $^{10}$C
``white'' stars are always associated with $^9$B$_{g.s.}$ decays. It is
possible that a Li nucleus, which is manifested weakly
in the dissociation of $^{10}$B (see Table 1), is also present
in $^{10}$B primarily in its ``dissolved'' form and produces a
nonresonance contribution to the $\Theta_{2\textrm{He}}$ distribution.

The study of $^{10}$B allows one to trace the evolution
from the cluster-type nuclear structure to the shelltype
one and requires the data on relativistic dissociation
of $^6$Li and $^9$Be nuclei and the identification of H
fragments in the 2He + H channel. A detailed understanding
of the coherent dissociation of $^{10}$B serves, in
turn, as the basis for interpreting the structure of the
next isotope ($^{11}$C).

\end{document}